\documentclass[aps,prd,twocolumn,groupedaddress]{revtex4}

  \usepackage{subfigure} 
\usepackage{graphicx}
\usepackage{epsf}

\newcommand{\sm}[1]{\mbox{{\scriptsize #1}}}

\newcommand{\bef}{\begin{figure}}
\newcommand{\eef}{\end{figure}}

\def\eps@scaling{0.95}

\def\showone#1{
  \centering
  \leavevmode
  \epsfxsize=\eps@scaling\linewidth
  \epsfbox{#1.eps}
}

\def\showtwover#1#2{
  \centering
  \leavevmode
  \epsfxsize=\eps@scaling\linewidth
  \epsfbox{#1.eps} \hfil
  \epsfxsize=\eps@scaling\linewidth
  \epsfbox{#2.eps}
}

\def\showthreeover#1#2#3{
  \centering
  \leavevmode
  \epsfxsize=\eps@scaling\linewidth
  \epsfbox{#1.eps} \hfil
  \epsfxsize=\eps@scaling\linewidth
  \epsfbox{#2.eps} \hfil
  \epsfxsize=\eps@scaling\linewidth
  \epsfbox{#3.eps}
}

\def\showfourover#1#2#3#4{
  \centering
  \leavevmode
  \epsfxsize=\eps@scaling\linewidth
  \epsfbox{#1.eps} \hfil
  \epsfxsize=\eps@scaling\linewidth
  \epsfbox{#2.eps} \hfil
  \epsfxsize=\eps@scaling\linewidth
  \epsfbox{#3.eps} \hfil
  \epsfxsize=\eps@scaling\linewidth
  \epsfbox{#4.eps}
}

\def\epstwo@scaling{0.46}

\def\showtwo#1#2{
  \centering
  \leavevmode
  \epsfxsize=\epstwo@scaling\linewidth
  \epsfbox{#1.eps} 
  \epsfxsize=\epstwo@scaling\linewidth
  \epsfbox{#2.eps}
}

\def\epsthree@scaling{0.28}
\def\showthree#1#2#3{
  \centering
  \leavevmode
  \epsfysize=\epsthree@scaling\textwidth 
  \epsfbox{#1.eps} 
  \epsfysize=\epsthree@scaling\textwidth 
  \epsfbox{#2.eps}
  \epsfysize=\epsthree@scaling\textwidth 
  \epsfbox{#3.eps}
}

\def\epstwo@scaling{0.44}

\def\showfour#1#2#3#4{
  \centering
  \leavevmode
  \epsfxsize=\epstwo@scaling\linewidth
  \epsfbox{#1.eps} \hfil
  \epsfxsize=\epstwo@scaling\linewidth
  \epsfbox{#2.eps} \hfil
  \epsfxsize=\epstwo@scaling\linewidth
  \epsfbox{#3.eps} \hfil
  \epsfxsize=\epstwo@scaling\linewidth
  \epsfbox{#4.eps}
}

\def\showsix#1#2#3#4#5#6{
  \centering
  \leavevmode
  \epsfxsize=\epstwo@scaling\linewidth
  \epsfbox{#1.eps} \hfil
  \epsfxsize=\epstwo@scaling\linewidth
  \epsfbox{#2.eps} \hfil
  \epsfxsize=\epstwo@scaling\linewidth
  \epsfbox{#3.eps} \hfil
  \epsfxsize=\epstwo@scaling\linewidth
  \epsfbox{#4.eps} \hfil
  \epsfxsize=\epstwo@scaling\linewidth
  \epsfbox{#5.eps} \hfil
  \epsfxsize=\epstwo@scaling\linewidth
  \epsfbox{#6.eps}
}

\newcommand{\befone}{
  \begin{figure*}
  \centering
  \begin{minipage}{\textwidth}
  }
\newcommand{\eefone}{\end{minipage}\end{figure*}}

\def\jnl@style#1{{\rmfamily#1}}%
\def\jref@jnl#1{{\jnl@style#1}}%

\newcommand\aj{\jref@jnl{AJ}}%
\newcommand\araa{\jref@jnl{ARA\&A}}%
\newcommand\apjl{\jref@jnl{ApJ}}%
\newcommand\apjs{\jref@jnl{ApJS}}%
\newcommand\apss{\jref@jnl{Ap\&SS}}%
\newcommand\aap{\jref@jnl{A\&A}}%
\newcommand\aapr{\jref@jnl{A\&A~Rev.}}%
\newcommand\aaps{\jref@jnl{A\&AS}}%
\newcommand\azh{\jref@jnl{AZh}}%
\newcommand\baas{\jref@jnl{BAAS}}%
\newcommand\jrasc{\jref@jnl{JRASC}}%
\newcommand\memras{\jref@jnl{MmRAS}}%
\newcommand\mnras{\jref@jnl{MNRAS}}%
\newcommand\pasp{\jref@jnl{PASP}}%
\newcommand\pasj{\jref@jnl{PASJ}}%
\newcommand\qjras{\jref@jnl{QJRAS}}%
\newcommand\skytel{\jref@jnl{S\&T}}%
\newcommand\solphys{\jref@jnl{Sol.~Phys.}}%
\newcommand\sovast{\jref@jnl{Soviet~Ast.}}%
\newcommand\ssr{\jref@jnl{Space~Sci.~Rev.}}%
\newcommand\zap{\jref@jnl{ZAp}}%
\newcommand\iaucirc{\jref@jnl{IAU~Circ.}}%
\newcommand\aplett{\jref@jnl{Astrophys.~Lett.}}%
\newcommand\apspr{\jref@jnl{Astrophys.~Space~Phys.~Res.}}%
\newcommand\bain{\jref@jnl{Bull.~Astron.~Inst.~Netherlands}}%
\newcommand\fcp{\jref@jnl{Fund.~Cosmic~Phys.}}%
\newcommand\gca{\jref@jnl{Geochim.~Cosmochim.~Acta}}%
\newcommand\grl{\jref@jnl{Geophys.~Res.~Lett.}}%
\newcommand\jgr{\jref@jnl{J.~Geophys.~Res.}}%
\newcommand\jqsrt{\jref@jnl{J.~Quant.~Spec.~Radiat.~Transf.}}%
\newcommand\memsai{\jref@jnl{Mem.~Soc.~Astron.~Italiana}}%
\newcommand\nphysa{\jref@jnl{Nucl.~Phys.~A}}%
\newcommand\physrep{\jref@jnl{Phys.~Rep.}}%
\newcommand\physscr{\jref@jnl{Phys.~Scr}}%
\newcommand\planss{\jref@jnl{Planet.~Space~Sci.}}%
\newcommand\procspie{\jref@jnl{Proc.~SPIE}}%

\begin{document}

\title{Cosmic constraints rule out {s}-wave annihilation of light dark matter}


\author{Dominik R. G. Schleicher, Simon C. O. Glover, Robi Banerjee, Ralf S. Klessen}
\email{dschleic@ita.uni-heidelberg.de}
\affiliation{Zentrum f\"ur Astronomie der Universit\"at Heidelberg, \\Institut f\"ur Theoretische Astrophysik,\\ Albert-Ueberle-Str.~2,\\ D-69120 Heidelberg,\\ Germany}

\begin{abstract}
Light dark matter annihilating into electron-positron pairs emits a significant amount of internal bremsstrahlung that may contribute to the cosmic gamma-ray background. The amount of emitted gamma-rays depends on the dark matter clumping factor. {Recent calculations indicate that this value should be of order $10^6-10^7$. That allows us to calculate the expected gamma-ray background contribution from dark matter annihilation. We find that the light dark matter model can be ruled out if a constant thermally-averaged cross section is assumed ({s}-wave annihilation). For more massive dark matter candidates like neutralinos, however, cosmic constraints are weaker.}
\end{abstract}


\maketitle

\section{Introduction}
{Recent observations of the Galactic center in different frequencies have provided increasing motivation for theoretical models that consider dark matter annihilation and/or decay. One finds an excess of GeV photons \citep{deBoer05}, of microwave photons \citep{Hooper07}, of positrons \citep{Cirelli08} and of MeV photons \citep{Jean06, Weidenspointner06}{, which correlates with the Galactic bulge instead of the disk and cannot be attributed to single sources. It is therefore controversial whether it can be explained by conventional astrophysical sources} alone or if dark matter annihilation models are required{. It is not clear whether these different phenomena are related. Hence, their interpretation is still under discussion \citep{deBoer08}.} For this reason, it is particularly interesting to also consider constraints which are independent of the Galactic center observations. 
}

{Observations in the MeV energy range favor models based on light dark matter. These models suggest that light dark matter particles with masses of {$1-100$~MeV} annihilate into electron-positron pairs \citep{BoehmHooper04}. In such a scenario, the dark matter mass needs to be larger than $511$~keV {to be able to produce electron-positron pairs through dark matter annihilation}. {It should be smaller than $100$~MeV, as otherwise {pion final states that produce too many gamma rays would be possible} \citep{BoehmHooper04}.} Electromagnetic radiative corrections to the annihilation process require the emission of internal bremsstrahlung \citep{Beacom05}. {It has also been proposed that such internal bremsstrahlung emission might explain the observed gamma-ray background in the $10-20$~MeV range \citep{Ahn05b, Ahn05a} for dark matter masses of $\sim20$~MeV, though these models appear less favorable in light of stronger upper limits on the dark matter particle mass \citep{Beacom06, Sizun06}. Supernovae data require dark matter particle masses larger than $10$~MeV, although this limit depends on assumptions made regarding the scattering cross section between dark matter particles and neutrinos \citep{Fayet06}.
}

{Calculations regarding the annihilation of light dark matter in the Milky Way show that predicted and observed fluxes are only in agreement for {p}-wave annihilation models \citep{BoehmHooper04}. In addition, dark matter models can be constrained by the cosmic backgrounds \citep{Ahn05c, Ando05, Rasera06, Yuksel07, Mack08, SchleicherBanerjee08c}. Such constraints provide a highly complementary approach {based on the observed cosmic gamma-ray background}. In this letter, we show that such constraints provide a strong independent confirmation that {s}-wave annihilation of light dark matter is ruled out. }

%

\section{Assumptions}
 {To explain the observed abundance of dark matter in the universe requires a thermally-averaged annihilation cross-section of \citep{Boehm04, Drees93}
\begin{equation}
\langle \sigma v\rangle\sim3\times10^{-26}\ \mathrm{cm^3}\ \mathrm{s}^{-1}.\label{cross}
\end{equation}
 In the mass range considered here, this may vary only by $10\%$ \citep{Ahn05a, Boehm04}. Such a cross section is in agreement with conservative constraints derived from gamma-ray observations of the Milky Way, Andromeda (M31) and the cosmic background \citep{Mack08}. We adopt Eq. (\ref{cross}) in this study.}

The overall intensity of annihilation radiation, however, depends sensitively on the dark matter clumping factor. This quantity, defined as $C(z)=\langle \rho_{\sm{DM}}^2(z) \rangle/\langle \rho_{\sm{DM}}(z)\rangle^2,$ has been subject to several studies \citep{Ahn05c, Ando05, Chuzhoy08, Cumberbatch08} and turns out to be highly uncertain. For example, a very detailed study by \citet{Cumberbatch08} computed clumping factors for three commonly-adopted dark matter halo profiles: Navarro-Frenk-White (NFW) \citep{Navarro97}, Moore \citep{Moore99} and Burkert profiles \citep{Burkert95}. Depending on model assumptions, the Burkert profiles, which have a flat central core, yield clumping factors of the order $10^5-3\times10^6$, the Moore profiles, which have a steep central cusp, yield factors in the range $10^7-10^{11}$, and intermediate NFW profiles yield factors of $10^6-3\times10^9$ at redshift zero. {The large variation in values for a given profile is due to the dependence of the clumping factor on further properties like the amount of substructure and typical halo concentration parameter. There has been a long controversy between different theoretical and observational studies, with observations typically favoring shallow profiles and theoretical investigations favoring power-law behavior with central slopes of the order $-1$ \citep{Flores94, Burkert95, Fukushige97, Fukushige03, Moore99, Salucci00, Ghigna00, Subramanian00, Taylor01, Jing00, Jing02, Klypin01, Blok02, Ricotti03, Gentile04, Tasitsiomi04, Hoekstra04, Broadhurst05}. More recently, there appears some convergence towards the Einasto profile \citep{Einasto65} with a slightly shallower slope of $\sim-0.8$ \citep{Graham06, Springel08a, Navarro08}. High-resolution simulations further indicate that substructure will not provide a strong contribution to the amount of dark matter annihilation \citep{Springel08b}{, though more realistic simulations including baryonic physics will be needed to finally resolve this issue}. As the slope of the Einasto profile is very close to the NFW, clumping factors should be in the same range, and certainly above the flat Burkert profile. This favors clumping factors in the range $10^6-10^7$, and as a firm lower limit, we adopt $10^5$ at redshift zero, which is the lowest value found for calculations with the Burkert profile, and a factor of $30$ below the lowest value for the NFW profile \citep{Cumberbatch08}. {We note that some works in the literature adopt an even smaller minimal clumping factor of $2\times10^4$ \citep{Ando05, Yuksel07, Mack08}, corresponding to the Kravtsov profile \citep{Kravtsov98}. Such a choice would leave our conclusions unchanged.}
}

\section{Formalism}\label{formalism}
The gamma-ray background intensity is given as \citep{Ahn05a} 
\begin{equation}
I_\nu=\frac{c}{4\pi}\int \frac{dz P_\nu([1+z]\nu,z)}{H(z)(1+z)^4},\label{int}
\end{equation}
where $c$ is the speed of light, $z$ the redshift, $P_\nu$ the proper volume emissivity and $H(z)$ the Hubble function. The proper volume emissivity is given as
\begin{equation}
P_\nu=\frac{1}{2}h\nu\langle\sigma v\rangle C_{\sm{clump}} n_{\sm{DM}}^2 \left[\frac{4\alpha}{\pi}\frac{g(\nu)}{\nu} \right],
\end{equation}
where $h$ is Planck's constant, $C_{\sm{clump}}$ is the dark matter clumping factor, $n_{\sm{DM}}$ the number density of dark matter particles, $\alpha\sim1/137$ the fine-structure constant and $g(\nu)$ is a dimensionless spectral function, defined as
\begin{equation}
g(\nu)=\frac{1}{4}\left(\ln\frac{\tilde{s}}{m_e^2}-1 \right)\left[1+\left( \frac{\tilde{s}}{4m_{\sm{DM}}^2} \right)^2 \right],
\end{equation}
with $\tilde{s}=4m_{\sm{DM}}(m_{\sm{DM}}-h\nu)$. The number density of dark matter particles is calculated from the mass density $\rho_{\sm{DM}}=m_{\sm{DM}} n_{\sm{DM}}$, which is highly constrained from WMAP observations \citep{Komatsu08}. As the thermally-averaged cross section is fixed by Eq.~(\ref{cross}), the predicted gamma-ray background depends only on the assumed dark matter particle mass $m_{\sm{DM}}$. { Between redshifts $10$ and zero, which yield the dominant contribution to the predicted gamma-ray background, the clumping factors are well described by a power-law 
\begin{equation}
C_{\sm{clump}}=C_0 \left(1+z\right)^{-\beta}
\end{equation}
with $\beta\sim3$ and $C_0$ being the clumping factor at redshift zero. As discussed above, the value of $C_0$ should be in the range $10^6-10^7$, and certainly above $10^5$. If the dark matter density in virialized halos can be assumed constant, the clumping factor rises as $(1+z)^{-3}$ towards lower redshift \citep{Chuzhoy08}. The evolution is however less steep if substructure in the halo is neglected \citep{Ahn05a,Ahn05c}. Such a case would strengthen the constraint given below, as even more radiation will be emitted at high redshift. }

\section{Results}\label{results}
For different dark matter masses, we calculate{ which clumping factor would be required to exceed the observed gamma-ray background due to emission of internal bremsstrahlung}. For illustrative purposes, we give some examples in Fig.~\ref{examples}. In Fig.~\ref{clump}, we show the constraint for different dark matter masses larger than $511$~keV. {Further limits can be calculated from $511$~keV line emission. {As a conservative choice, we assume that the electron-positron-pairs annihilate via positronium formation as observed in the Milky Way \citep{Kinzer01}, such that only $25\%$ of all annihilations lead to the emission of $511$~keV photons.} For masses of $7$~MeV, one finds a maximum permitted clumping factor of $C_0=10^5$ \citep{SchleicherBanerjee08c}. As this emission always peaks at the same frequency, one can obtain the constraint for other particle masses by requiring that $C_0/m_{\sm{DM}}^2=\mathrm{const}$. For dark matter masses larger than $11$~MeV, the constraint from internal bremsstrahlung is most stringent. The limit from $511$ keV emission is only important for dark matter masses below $11$ MeV. For masses below the upper limit from Galactic center observations, the constraint from $511$~keV emission is strongest.}
\bef
\showone{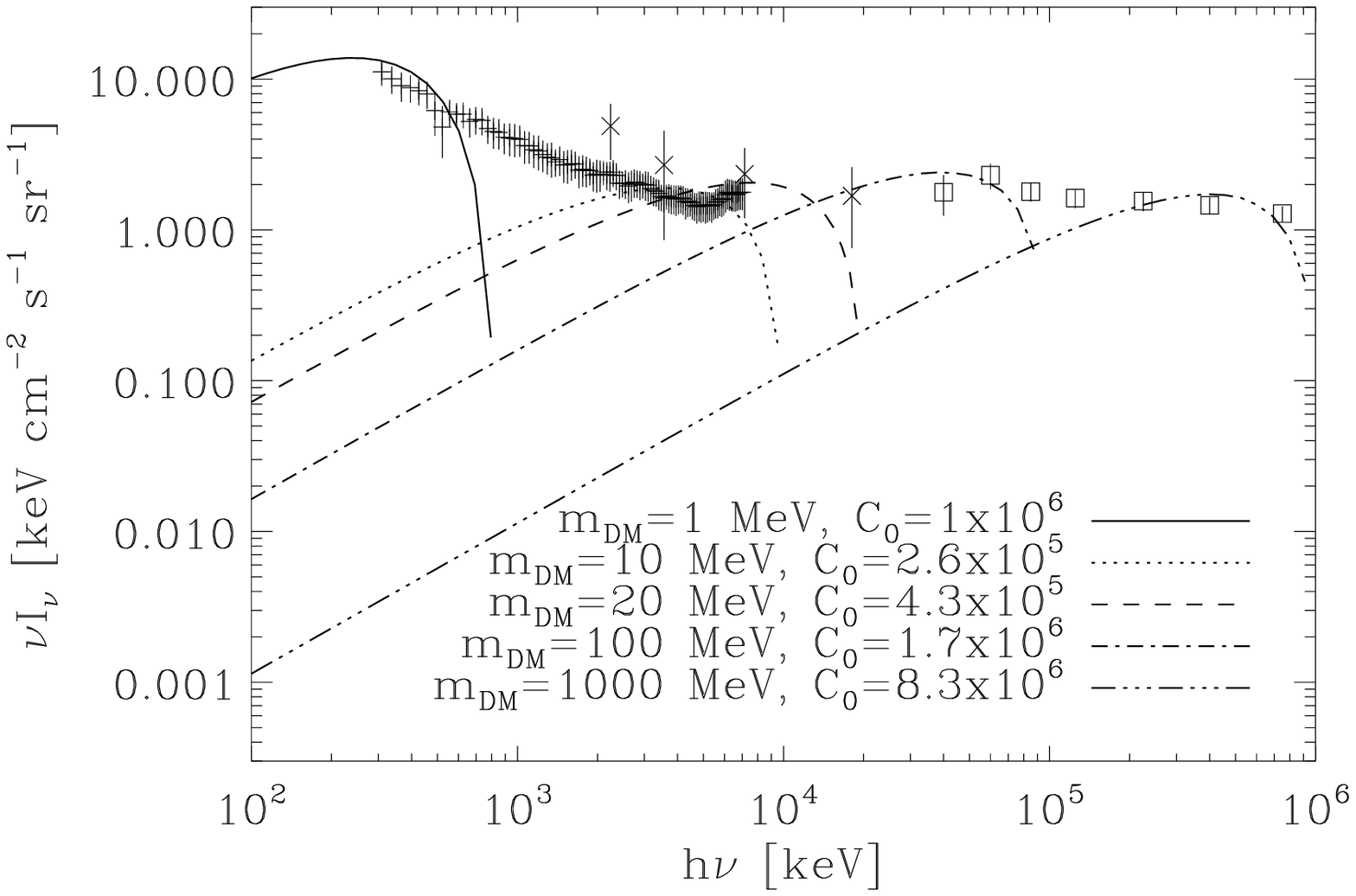}
\caption{The predicted gamma-ray background due to internal bremsstrahlung emission for different dark matter particle masses. In every case, we adopted a clumping factor that yields the maximum allowed background. We compare with the observed gamma-ray background from COMPTEL (crosses) \citep{Kappadath96}, SMM (plusses) \citep{Watanabe99} and EGRET (squares) \citep{Strong04}.}
\label{examples}
\eef

\bef
\showone{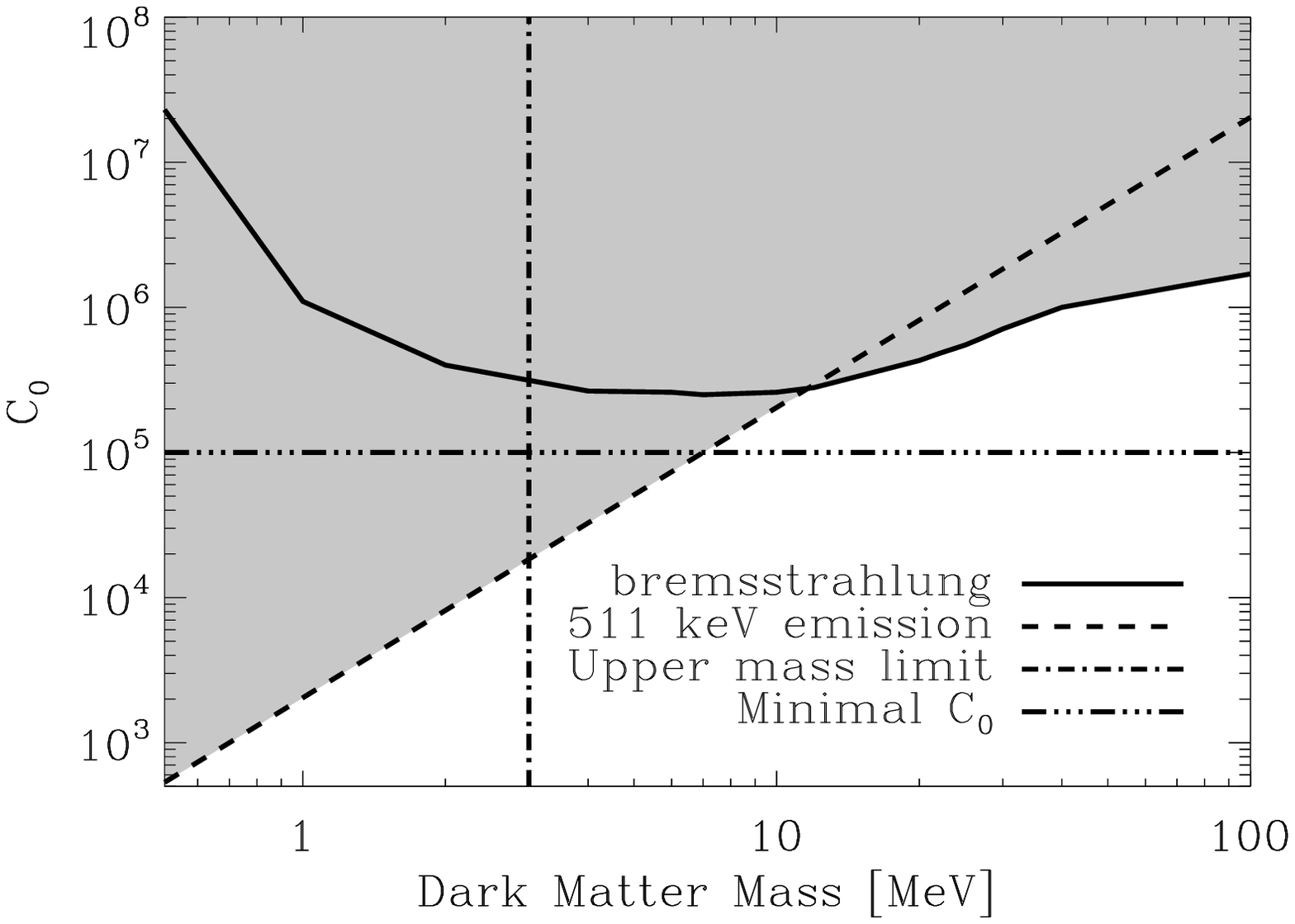}
\caption{The constraint on the clumping factor $C_0$ at redshift zero due to internal bremsstrahlung emission. For comparison, we show the weaker constraint due to $511$ keV emission. The forbidden region is shaded.{ We also show the upper mass limit from Galactic center observations calculated by \citet{Beacom06} as well as the minimal clumping factor $C_0=10^5$. The combination of these constraints shows that light dark matter models with {s}-wave annihilation are ruled out.}}
\label{clump}
\eef

\section{Implications and discussion}

{As shown by \citet{Beacom06}, light dark matter scenarios that explain observations of the Galactic center require a mass less than $3$~MeV. {This results from a comparison between $511$~keV emission and higher-energy gamma-rays and does not require assumptions regarding the dark matter profile. For low} masses, clumping factors $C_0$ in the range $10^6-10^7$, which we would expect for the currently-favored Einasto profiles, are excluded due to the gamma-ray background constraints. Even if we adopted clumping factors corresponding to a flat Burket profile with little substructure, corresponding to values $C_0\ge10^5$ \citep{Cumberbatch08}, this would violate the constraints from the gamma-ray background. This shows that the light dark matter model can no longer be maintained with a constant thermally-averaged annihilation cross section if constraints from the cosmic background {are combined with the upper mass limits from the spectral constraints{, thus providing an independent confirmation that s-wave annihilation of light dark matter can be ruled out.}}


{These conclusions hold even if we adopted the very conservative upper mass limit of \citet{Sizun06}, which is $7.5$~MeV if the ISM in the Galactic center would be strongly ionized. This appears unlikely and provides a very strong upper limit. In addition, more recent INTEGRAL observations favor an even smaller emission region, which increases the inflight annihilation intensity by a factor of $1.7$ and pushes the allowed maximum positron injection energy even further down (John Beacom, private communication). }

{Still feasible are models based on {p}-wave annihilation, in which the annihilation cross section becomes much smaller at late times when the velocities of dark matter particles are no longer relativistic \citep{BoehmFayet04}. It is also interesting to note that the gamma-ray background at $10-20$~MeV might be explained by a power-law component of non-thermal electrons in active galactic nuclei (AGN) \citep{Inoue08}. It is however unclear whether a sufficient number of non-thermal electrons is actually available. We expect that the FERMI satellite \footnote{http://www.nasa.gov/mission\_pages/GLAST/science/index.html} will shed more light on such questions, as the anisotropic distribution of the gamma-ray background may allow one to distinguish between astrophysical sources and dark matter annihilation \citep{Ando07}.} 

{We finally note that we have checked if similar constraints apply for more massive dark matter candidates, based on the recent analysis of \citet{Mack08} and \citet{Yuksel07}. For such models, however, clumping factors in the range $10^7-10^9$ are still feasible and well within the allowed parameter space. {The upper limits derived for light dark matter do not apply here, as the annihilation products may be different.}
}

{
We finally note that if the clumpiness of dark matter was high already at early times, annihilation of light dark matter may provide a significant contribution to the observed reionization optical depth \citep{SchleicherBanerjee08a, Chuzhoy08, Natarajan08}. The detailed effects of such annihilations on the IGM have been explored by \citet{Ripamonti07}, and consequences for $21$~cm observations have been explored by \citet{Furlanetto06, Chuzhoy08}.}

\begin{acknowledgments}
We thank Kyungjin Ahn for interesting discussions on dark matter annihilation and the gamma-ray background, Duane Gruber for providing the HEAO and Comptel-data and Ken Watanabe for providing the SMM data. {We further thank John Beacom, Pierre Fayet, Marco Spaans and Volker Springel for interesting discussions on a previous version of this manuscript.} DS thanks the Heidelberg Graduate School of Fundamental Physics (HGSFP) and the LGFG for financial support. The HGSFP is funded by the Excellence Initiative of the German Government (grant number GSC 129/1). SCOG acknowleges support from the German Science Foundation (DFG) under grant Kl 1358/4. RB is funded by the Emmy-Noether grant (DFG) BA 3607/1.  RSK thanks for support from the Emmy Noether grant KL 1358/1. All authors also acknowledge subsidies from the DFG SFB 439 {\em Galaxies in the Early Universe}. {We further thank the anonymous referees for valuable remarks on the topic.}
\end{acknowledgments}

\clearpage




\clearpage

\end{document}